\documentclass[twoside,11pt,english]{article}

\usepackage[T1]{fontenc}

\usepackage{palatino}
\usepackage{graphicx}
\usepackage{amsmath}
\usepackage{amssymb}

\newcommand{\msol}{$M_\odot$}
\newcommand{\mmsol}{M_\odot}

\begin{document}

\vspace*{-1.8cm}
\begin{flushright}
{\bf LAL 04-70}\\
\vspace*{0.1cm}
{July 2004}
\end{flushright}
\vspace*{0.5cm}

\begin{center}
{\LARGE\bf EROS: a Galactic Microlensing Odyssey}
\vspace*{0.5cm} 

{\Large {\bf R. Ansari} (EROS collaboration)}

\vspace*{0.5cm}
{\bf\large Laboratoire de l'Acc\'el\'erateur Lin\'eaire,}\\
IN2P3-CNRS et Universit\'e de Paris-Sud, B\^at.\ 200, BP 34, 
F-91898 Orsay Cedex, France 
\end{center}

\begin{abstract}
The EROS microlensing survey has monitored nearly 100 million stars for seven 
years, to search for halo brown dwarfs and compact objects in the Galactic 
disk. In this paper, we review the various EROS observation programs and
the corresponding microlensing search results. In particular, based on 
LMC and SMC observations, EROS excludes a major contribution from 
compact objects with masses in the range $2. \times 10^{-7} - 1 \, M_{\odot}$ to the 
Galactic halo. Less than 25\% of the standard halo mass can be made of
such objects, while the EROS measured optical depths toward the Galactic 
spiral arms ($<\tau_{SA}> \simeq 0.43 \times 10^{-6}$) and Galactic bulge 
($<\tau_{GC}> \simeq 0.93 \times 10^{-6}$) are compatible with Galactic model
predictions if the contribution from an elongated bar in the centre
is taken into account.
\end{abstract}
\section{Introduction}
EROS (Exp\'erience de Recherche d'Objets Sombres) odyssey started in 1990,
when J. Rich and M. Spiro initiated one of the first microlensing 
surveys, following B. Paczynski suggestion, to search for compact 
dark halo objects. 
\par
The EROS project, a mostly French collaboration involving particle 
physicists and astrophysicists from DAPNIA, IN2P3 and INSU started
LMC and SMC observations at the end of 1990, using photographic 
plates taken with the ESO Schmidt telescope at La Silla observatory
(Chile). Nearly a year later, a dedicated wide field CCD camera 
mounted on a 40 cm telescope started operations at La Silla.
The Schmidt photographic plates where used to search for
long duration ($\sim$ days) microlensing events, while the CCD-T40
images provided sensitivity for short duration events. 
\par
in 1994, a more ambitious program was approved by the EROS team and 
the funding agencies in order to disentangle contributions 
from different Galactic components (disk, bar, halo) to 
the optical depths.
The enlarged EROS collaboration, with Danish and Chilean participation started then
 the commissioning of EROS 2- MARLY dedicated wide field 
 telescope and camera system. This instrument and EROS 2 observation 
 programs are described in section 3. Section 2 presents the basics of 
 microlensing and compact massive halo object  (MACHO) search. 
 In section 4, I describe the EROS 2 LMC and SMC observation 
 programs and microlensing search results.   The EROS observation 
 programs toward the Galactic spiral arms and the bulge, as well as the 
 observed optical depths in these directions are presented in section 5.
 \section{Microlensing basics}
 It is now rather well established that a large fraction of matter in the 
 universe is in a unknown non luminous form, other than stars, gas and
dust. The dynamics of galaxies, in particular the rotation velocities of stars 
and gas in spiral galaxies, including the Milky way cannot be explained by the 
distribution of the visible matter and favours the presence of dark  haloes, 
with masses up to ten times the visible mass.
In addition, the baryonic matter density inferred from Big Bang Nucleosynthesis or
recent CMB anisotropy measurements is much larger that the visible matter density, 
implying that dark haloes could be baryonic.
\par
Brown dwarfs, sub stellar objects too light to burn hydrogen,
are one of the plausible candidate for baryonic halo dark matter. 
The massive compact halo objects are very often called MACHOs 
(Massive Compact Halo Objects).
The mass range extends from the evaporation limit ($10^{-7}$ \msol) 
to the ignition limit (0.08 \msol) for brown dwarfs, and extends well above
1 \msol for stellar remnants and black holes.
A detailed discussion of various
candidates for MACHOs can be found in \cite{carr94}, 
\cite{kerins94} and \cite{carr00}.
\par
Direct observation of MACHOs is nearly impossible. However, as
pointed out by Paczynski in 1986 \cite{paczyn86}, their presence
in the Galaxy may be revealed through their 
gravitational microlensing effect on background stars. 
\par
A massive object $L$, passing close enough to the line of sight OS
for a star, acts as a gravitational lens and induces a 
relativistic light deflection. Multiple images of the source 
are generally produced, well known in case of the lensing of distant galaxies
by foreground galaxies or clusters. In the case of Galactic MACHOs,
image separation is of the order of $10^{-3}$ arcsec. The two
images can not be resolved, and lensing leads then 
to an increase of the apparent brightness of the star. This 
light amplification is simply due to collection by the observer
of light in a larger solid angle from the source 
in the presence of the deflector.
\par
For a point like source ($S$) and deflector ($L$), and an observer at 
$O$, the The light amplification $A$ can be expressed as a function of the 
reduced impact parameter $u$ :
\begin{displaymath}
A = \frac{u^2+2}{u\sqrt{u^2+4}}  \hspace{0.5cm} ,  \hspace{0.5cm} 
u=\frac{D_\perp(OS,L)}{R_E} 
\end{displaymath}
and the Einstein ring radius $R_E$ in the deflector plane can be written 
as:
\begin{displaymath}
R_E^2 = \frac{4 G M_L D}{c^2}  \hspace{0.5cm} , \hspace{0.5cm} 
D = \frac{D_{OL} D_{LS}}{D_{OS}} 
\end{displaymath}
Due to the deflector movement relative to the line of
sight, the impact parameter u is time dependent, and 
hence is the observed light amplification. The lensing becomes
then an observable phenomenon.
\par
The amplification time scale is set by the time interval 
$t_E = R_E / v_\perp$ taken by the deflector to cross the Einstein
ring radius, where $v_\perp$ is the projected transverse velocity of the 
deflector relative the line of sight $OS$. 
The time varying reduced impact parameter $u(t)$
can then be written as a function of $u_0$, the reduced impact parameter at the 
time of closest approach $t_0$:
$$u^2(t) = u_0^2+\frac{(t-t_0)}{t_E}^2$$
For stars in the Large Magellanic Cloud, and 
deflectors in our Galaxy halo, with a maxwellian
velocity distribution at 200 km/s, the average duration 
is $\tau \simeq 70\ {\rm days} \times \sqrt{M/\mmsol} $.
\par
Measurable amplification occurs only for well aligned observer,
deflector and source systems ($A_{max} > 1.34$ for $u_0 < 1$). The 
lensing probability for a given amplification 
The optical depth corresponds to the lensing probability with
amplification $A>1.34$ and can be expressed as the 
fraction of the sky covered by MACHO's Einstein disks.
The Einstein disk surface is proportional to $R_E^2$, 
which is proportional to $M_L$, whereas the number of deflectors, 
for a given total mass is proportional to $M_L^{-1}$.
The optical depth $\tau$ is thus nearly independent of
the mass of the MACHOs. Typically, $\tau = 0.5\ 10^{-6}$
for stars in the LMC
and for a ``standard'' halo made of MACHOs
($M_{tot} \simeq 4.\ 10^{11} \mmsol$ up to 50 kpc). 
\par 
This very low lensing probability on one hand ($\sim 10^{-7}$), 
and the difficulty of distinguishing 
genuine microlensing events from the background
of variable stars make the search for MACHOs a 
challenging task. The intrinsic properties of microlensing
effect is used in background rejection, in particular the symmetry of
the light curve, its specific shape and achromaticity.
However, the light curve shape is altered when more subtle effect
such as the finite size or the parallax effects are taken into account.
A discussion of these effects can be found in \cite{derue99} and 
\cite{rahvar03}. In the case of binary lenses, the light curve
present spectacular and characteristic features when caustic crossing
occurs \cite{dominik99}.
\section{EROS-2 instrument and observation program}
\subsection{EROS2-MARLY instrument}
The MARLY telescope (D=1 m, f/5) has been specially refurbished and 
fully automated for the EROS-2 survey \cite{bauer97}. The telescope
optics allows simultaneous imaging in two wide pass-bands 
$V_{Eros}$ ($\lambda_{peak} \simeq$ 600 nm, $\Delta \lambda \sim$ 80 nm), 
$R_{Eros}$ ($\lambda_{peak} \simeq$ 760 nm, $\Delta \lambda \sim$ 80 nm) 
\cite{regnault00} over a one-square-degree field of vue. 
The two focal planes are 
equipped with CCD cameras, each made of a mosaic of 
8 $(2 \times 4)$ Loral $2048 \times 2048$ thick CCD's, covering 
a total field of $0.7^{\circ}$ (right ascension) 
$\times$ $1.4^{\circ}$ (declination). The pixel size is 0.6 arcsec, with 
a typical global seeing of <2 arcsec FWHM.
Each camera has more than 32 millions pixels, representing a total
of 128 megabytes of image data for each frame.
\par
\begin{figure}[h]
\begin{center}
\mbox{\includegraphics[width=10cm,height=7cm]{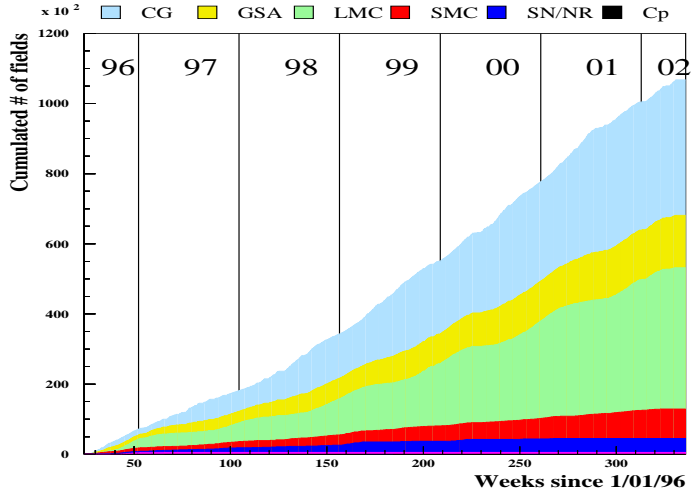} }
\caption{EROS cumulative image data distribution. CG: Galacatic Bulge, 
 GSA: Galactic Spiral Arms, LMC,SMC: Large and Small Magellanic Clouds, 
 SN/NR: Supernovae/proper motion, Cp: Cepheids}
\label{figdatavol}
\end{center}
\end{figure}

The  EROS2-MARLY telescope-camera system has been installed 
at La Silla observatory in Chile in spring 1996.
Regular operations were started in July 1996, and has been 
carried out up to February 2003. During these seven years, EROS
accumulated more than $\sim 2 \times 10^6$ $2K\times2K$ image frames,
representing $\sim 15$ terabytes of data. The cumulative image
data distribution is shown in figure \ref{figdatavol}.
\subsection{Scientific programs}
\par
EROS-2 was primarily designed and optimized to 
measure microlensing optical depths for different lines 
of sight through the Galactic disk and halo.
Systematic photometric surveys were carried out in several directions,
with typical sampling times of a few days. 
Around 80 fields are being monitored toward the Large Magellanic Cloud (LMC)
and 10 fields toward the SMC. A large area in direction of the 
bulge is also included in our survey (CG , $\sim$ 150 fields) as well 
as 29 fields in the Galactic plane, away from the bulge.
\par
Type I supernovae are being used as standard
candles to probe the Universe geometry. 
EROS has discovered around 60 supernovae in the period 
1997-1999. 25 have been spectroscopically identified, 
of which 20 are of type Ia. The EROS automated SN search has
been able to discover $\sim 1$ SN / 2 hours observing time, or
 $\sim 1$ SN / 10 - 20 deg$^2$ with an average redshift of 
 $<z> \sim 0.1$ \cite{regnault00}.
\par
A small fraction of the EROS telescope time 
 has been used to search for red and white dwarves
 in the solar neighbourhood  through proper motion measurements.
 The constraints from this survey can be found in \cite{goldman02}
%
\subsection{Data management and photometric pipeline}
The EROS data is stored at the IN2P3 computing centre (CC-IN2P3)
at Lyon (France) where a hierarchical storage system (HPSS)
\footnote{More information available from the CC-IN2P3 web site 
http://webcc.in2p3.fr/} is used.
Data files (Images, catalogs, lightcurves \ldots)
management is done using the Oracle relational database, with 
the help of the {\bf ErosDb} software written in Java and Tcl.
Task scheduling and monitoring for the photometric 
and lightcurve processing pipeline is also  handled by {\bf ErosDb}.
The {\bf PEIDA++} \footnote{PEIDA++ documentation is available from
http://www.lal.in2p3.fr/recherche/eros/PeidaDoc/index.html}
C++ class library is the basis for the different photometric, image 
processing and light curve analysis software modules.
The standard EROS photometric pipeline \cite{ansari96} 
performs PSF photometry for each of the images using 
a reference star catalog. This catalog contains star positions and 
reference flux obtained from a high signal to noise ratio reference 
image. The reference image is usually obtained 
by the coaddition of a set of 10 to 20 good quality images.
The main steps of processing for a given image frame are: the geometric 
alignment, PSF photometry with fixed positions and PSF parameters,
and photometric alignment.
The light curve data base is then updated with the new flux measurement
for each of the stars in the reference catalog.
\section{Magellanic clouds observations}
\subsection{SMC}
\begin{figure}[h]
\begin{center}
\mbox{\includegraphics[width=7cm]{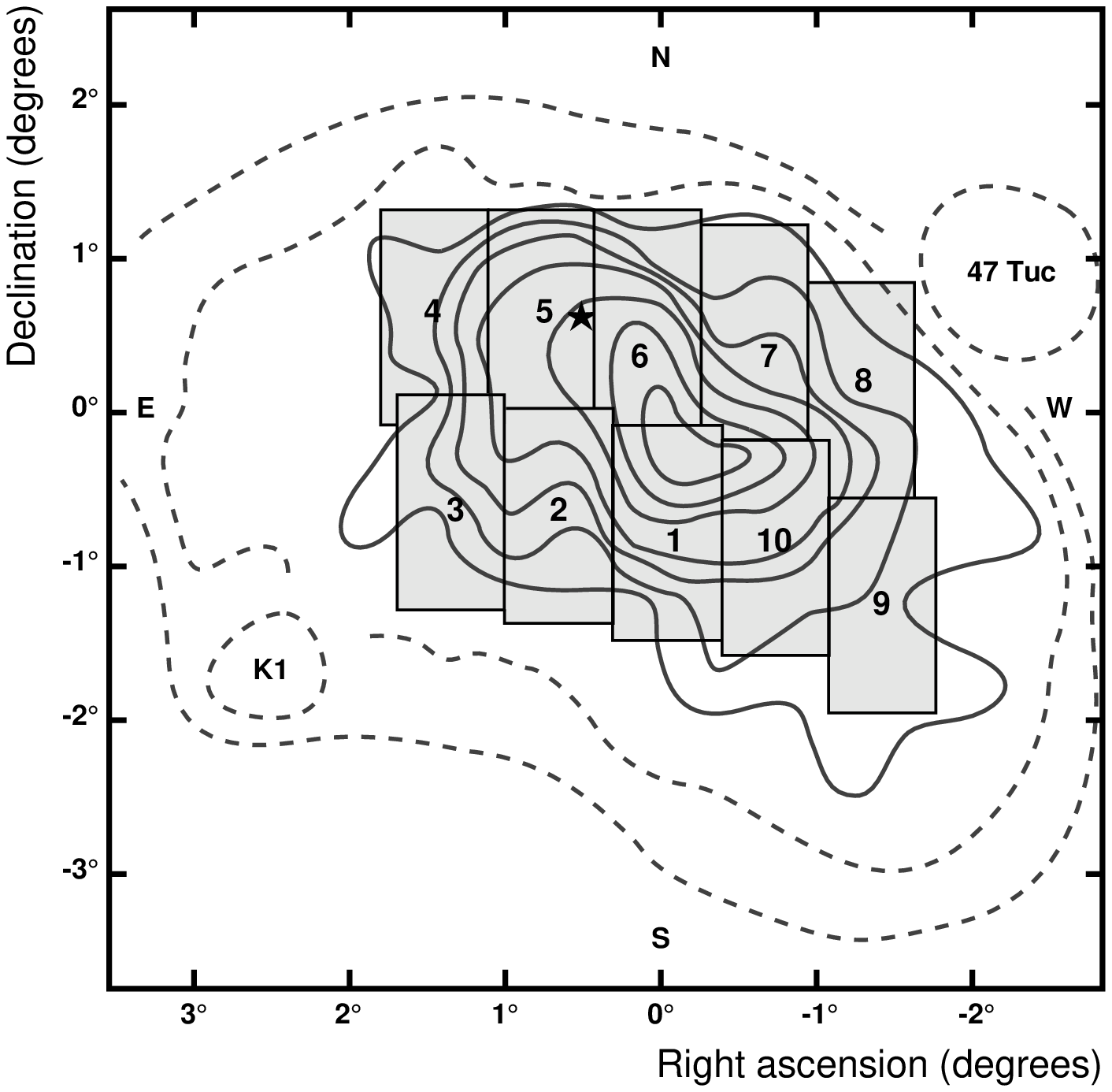}}
\mbox{\includegraphics[width=7cm,height=7cm]{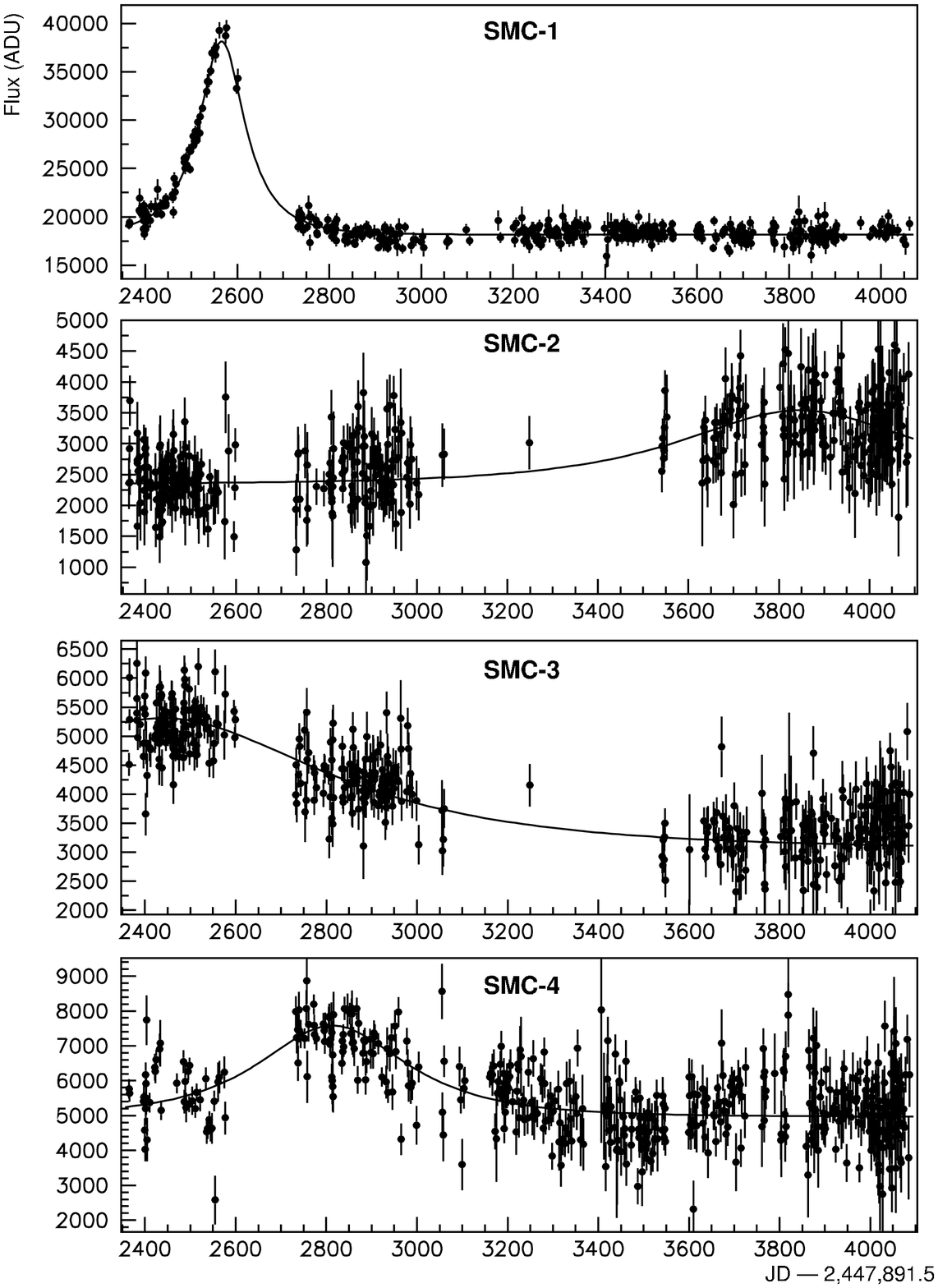}}
\caption{Left: EROS SMC fields. Right: Four SMC microlensing candidates}
\label{figsmc}
\end{center}
\end{figure}

Ten fields, representing $8.6 deg^2$ over the Small Magellanic Cloud
has been monitored by EROS, with exposure times ranging from 5 to 15 minutes.
Figure \ref{figsmc} (left) shows the field positions
on the sky. The latest published analysis correspond to five years SMC data
(July 1996 - March 2001) and $5.2 \times 10^6$ light curves with 
 $ \sim 400-500$  images in each of the two EROS pass bands (B,R)
\cite{smc5y}. 
Four microlensing (SMC-1,2,3,4) candidates has been found in this analysis. Their
light curves are shown on figure \ref{figsmc} (right).
The montecarlo estimated efficiencies for events with $u_0<1.5$ and normalised 
to the 5 years observation period reaches a maximum of $\sim 14\%$ 
for event time scales of $t_E \sim 100$ days, and drops to less than 3\% 
for time scales $t_E$ below 10 days or above 1300 days.
\par
SMC-1 has a duration $t_E \sim$ 101 days and corresponds probably to a deflector in SMC
itself. The three other candidates (SMC-2,3,4) are long duration events with 
$t_E \sim$ 390, 612, 243 days. These events are probably misidentified variable
stars or due to self-lensing within the cloud. Indeed, deflectors in the halo would need
to have supersolar masses to account for such durations.
\subsection{LMC}
EROS has monitored 89 fields covering the large magellanic cloud.
The published analysis has been carried out on 39 $\deg^2$ well
sampled fields, using data from the three first years. A total
of $25.5 \times 10^6$ light curves with 100-200 measurements
have been analysed, with selection efficiencies ranging from 
5\% ($t_E \sim 5$ days) to 20\% (($t_E \sim 100$ days).
The characteristics of the four microlensing candidates
are shown in table \ref{evtlmc}. Detailed description of
the two years and three years analysis can be found in 
\cite{lmc2y} and \cite{lmc3y}.
\begin{table}[h]
\begin{center}
\begin{tabular}{l|cccc}
\hline
 &$u_0$&$t_E$& $\chi^2/{\rm dof}$ & $V_J$ \\
\hline
 LMC-3 & $0.21$&$44$ & 219/143 & 22.4\\
\hline
LMC-5 & $0.58$&$24$ & 658/176 & 19.2\\
\hline
 LMC-6 & $0.38$&$36$ & 682/411 & 21.3\\
\hline
 LMC-7 & $0.23$&$33$ & 722/356 & 22.7\\
\hline
\end{tabular}
\caption{Characteristics of LMC microlensing candidates} 
\label{evtlmc}
\end{center}
\end{table}
\subsection{halo constraints}
\begin{figure}[h]
\begin{center}
\mbox{\includegraphics[width=12cm]{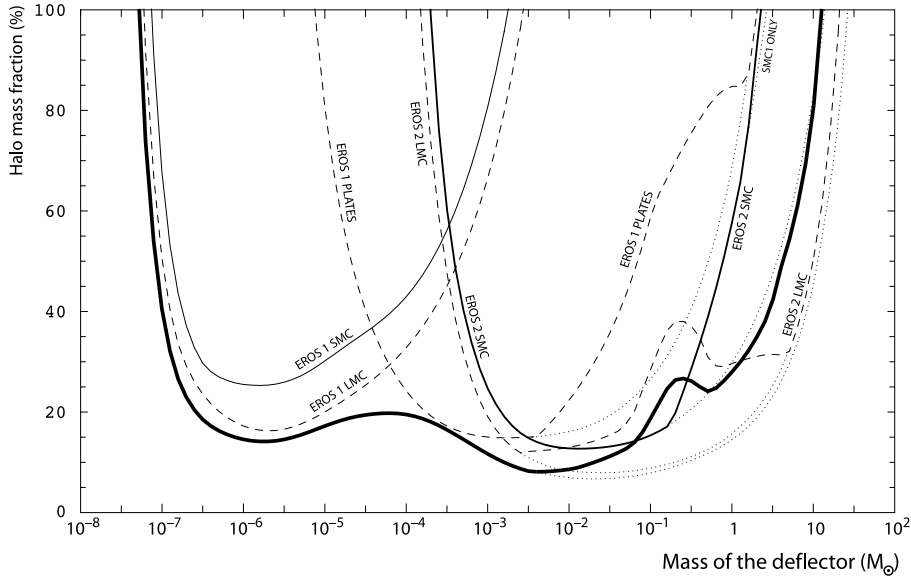}}
\end{center}
\caption{Exclusion diagram at 95\% C.L. for the standard halo
model ($4 \times 10^{11} M_\odot$ inside 50 kpc). The thick line
is the combined limit from five different EROS data sets and analysis
\cite{smc5y}
}
\label{limithalo}
\end{figure}

The combined EROS constraints on the contribution of 
compact objects to the galactic halo is shown in figure
\ref{limithalo}. massive compact objects with masses in the range
$2 \times 10^{-7} M_\odot$ and $1 M_\odot$ cannot represent more 
than 25\% of the halo mass, in the case of a standard spherical, isothermal 
Galactic halo, encompassing $4 \times 10^{11} M_\odot$ out to
50 kpc. These limits have been obtained in a conservative approach, 
where all the candidate microlensing events are attributed to 
deflectors in the halo, and taking into account the 
corresponding event durations.
\par
However, it should be noted that SMC and LMC microlensing candidates
have different event duration distribution and can hardly be due to the 
same lens population. In the case of a standard halo, the 
optical depth toward SMC would be slightly larger than 
toward LMC ($\tau_{SMC}\sim 1.4\tau_{LMC}$). Based on MACHO
optical depth and event duration distribution from MACHO, we 
would expect three short duration events ($<t_E> \sim 30$ days)
where none is observed.
\section{Galactic plane observations}
\subsection{Spiral arms (GSA) microlensing search}
\begin{figure}[h]
\mbox{\includegraphics[width=8cm,height=5cm]{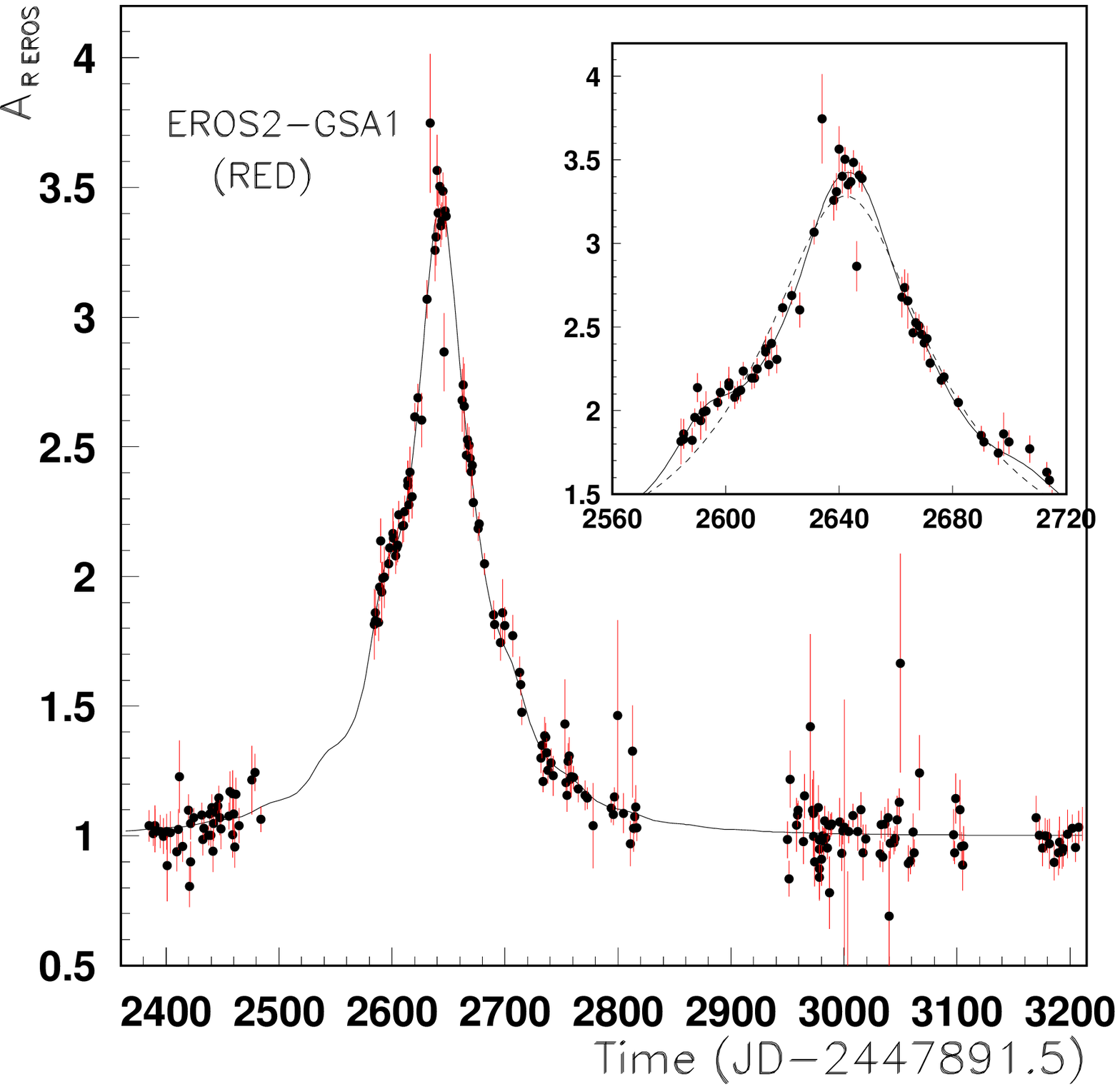}}
\mbox{\includegraphics[width=8cm,height=5cm]{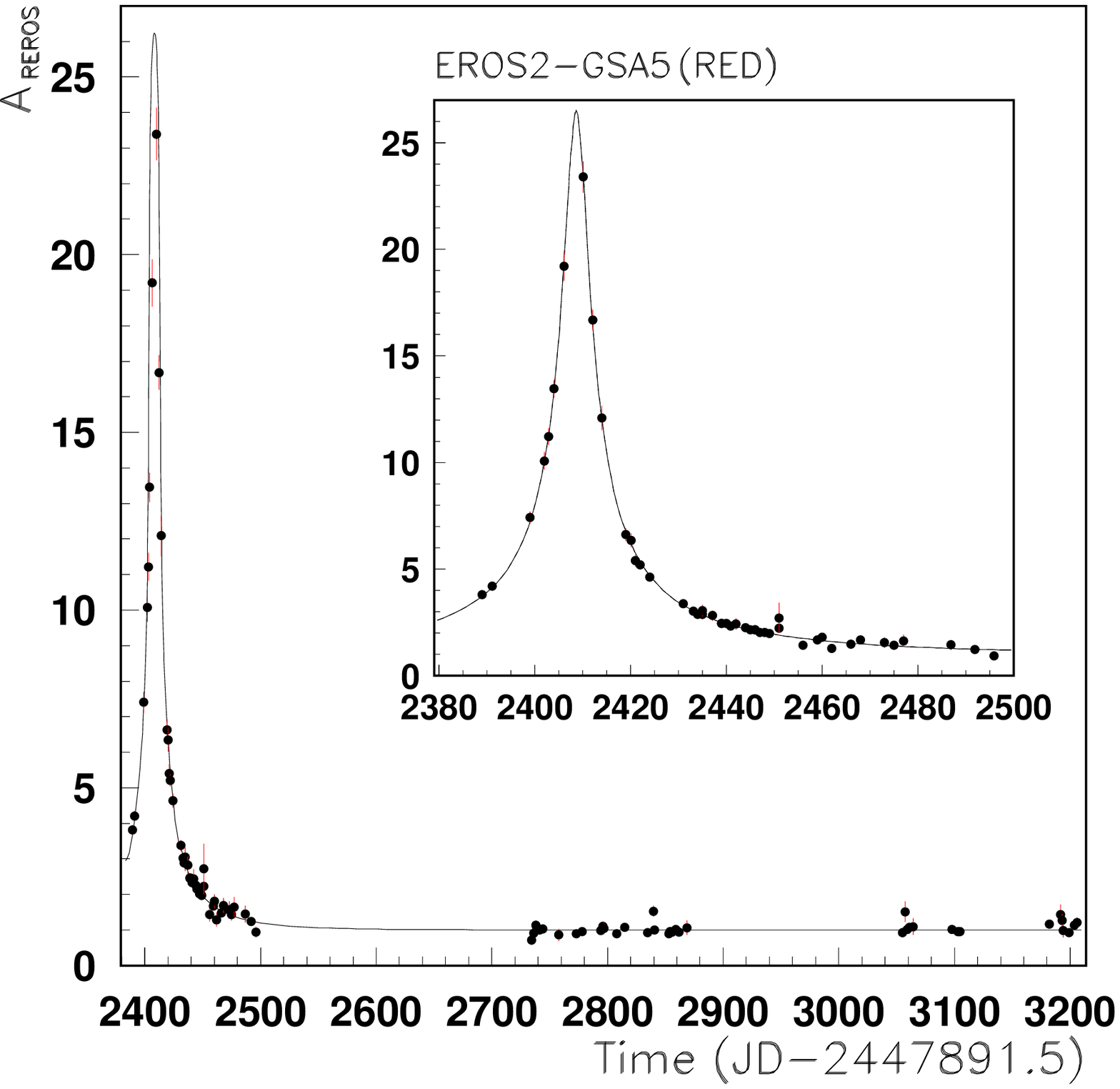}}
\caption{Left: GSA-1, interpreted as a microlensing on a binary source
star. Right: GSA-5, large amplification event. We have been able to 
obtain constraints on the deflector based on the absence of parallax 
and finite size effect. }
\label{lcgsa}
\end{figure}

The 29 Galactic plane fields (GSA) are grouped in four directions 
$(\beta_{Sct} , \gamma_{Sct}$, $\gamma_{Nor} , \theta_{Mus} )$ 
and cover a wide range of longitude.
The three year data set discussed here contains 9 million 
light curves : 2.1 towards $\beta_{Sct}$, 1.8 towards $\gamma_{Sct}$, 
3.0 towards $\gamma_{Nor}$ and 2.1 towards $\theta_{Mus}$.
The analysis corresponds
to data taken from  July 1996 to November 1998, 
except for $\theta_{Mus}$ which has been monitored since January 1997.
Detailed description of the analysis can 
be found in \cite{bs3y}.
Seven light curves satisfy all the selection criteria and are 
labelled GSA1 to 7. Five events have been found toward 
$\gamma Sct$ with durations $t_E \sim$ 6, 24, 38, 59 and 72 days and
two events with $t_E \sim$ 72 and 98 days toward $\gamma Nor$.
Two of the most interesting events (GSA-1 and GSA-5)
are shown in figure \ref{lcgsa}.
\par
We have computed the expected optical depths using a three component
model (bar, disk, halo) to represent deflector distribution in the Galaxy.
An average distance of $\sim 7$ kpc
has been used in this analysis. Figure 5 shows the 
expected optical depth up to 7 kpc at galactic latitude $b=-2.5^\circ$
as a function of the Galactic longitude,
for two sets of bar parameters. 
The corresponding values for the four targets are also indicated on
figure 5. 

\begin{figure}[h]
\begin{center}
\mbox{\includegraphics[width=12cm,height=7cm]{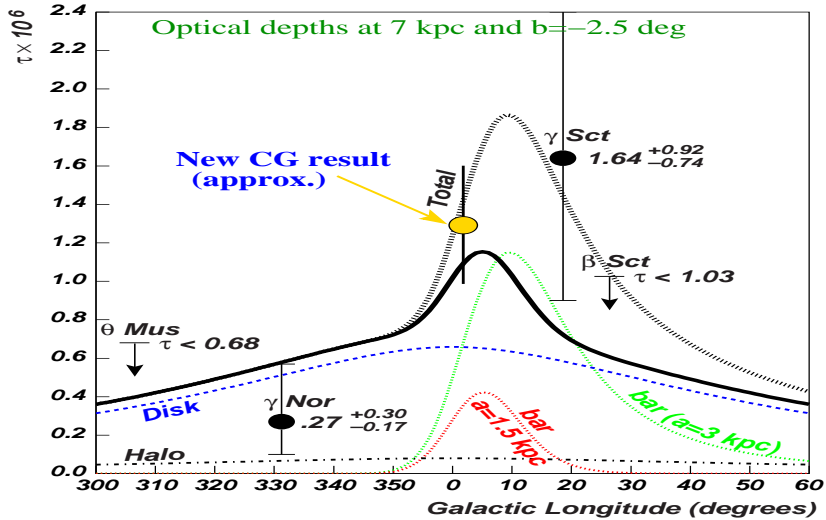}} 
\caption{Expected and observed optical depths toward different
targets in the Galactic disk. The indicated approximate CG result has 
been computed using the EROS measured optical depth, shifted in order 
to take into account the target distance (7 kpc) and galactic latitude 
($b=-2.5^\circ$)}
\end{center}
\label{taugsa}
\end{figure}

\par
We find an estimated optical depth averaged over the four directions  
$$ <\tau_{GSA}> = 0.43^{+0.24}_{-0.11} \times 10^{-6} $$
in agreement with expectations. However, as shown on figure 5,
optical depths or limits computed for each direction indicates a small
excess toward $\gamma_{Sct}$ direction (mainly due to short events), 
compared to other directions.
\subsection{galactic bulge fields}
2.3 million stars, observed on 15 best sampled Galactic bulge fields out of 82,
have been analysed, using the first three years data. 
Each light curve has a total of 200 to 400 measurement points.
Detailed description of the analysis can be found in \cite{bulge3y}.
The search for microlensing events has been restricted to clump red giant stars,
in order to minimize the blending effect and yielded a total of 16 
candidates, two of which are long duration events.
We find an optical depth 
$$\tau_{CG} = (0.94 \pm 0.29) \times 10^{-6}$$
well in agreement with model predictions.
Figure 6 shows one of the EROS bulge events, as well 
as optical depths measured by different microlensing surveys toward
the Galactic bulge.
\vspace*{0.5cm}
\begin{figure}[h]
\begin{center}
\mbox{\includegraphics[width=7cm,height=6cm,clip]{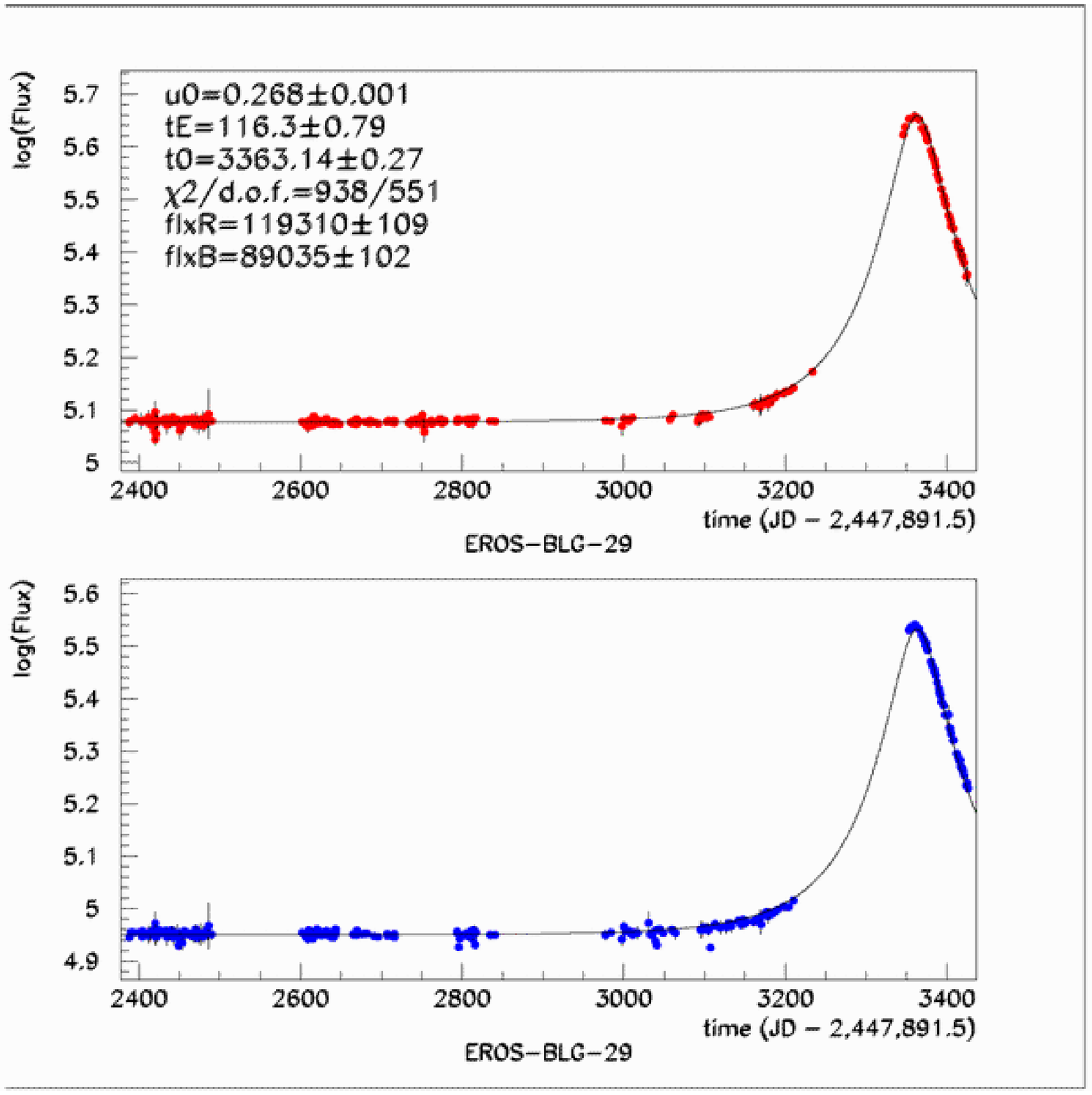}}
\mbox{\includegraphics[width=8cm,height=6cm,clip]{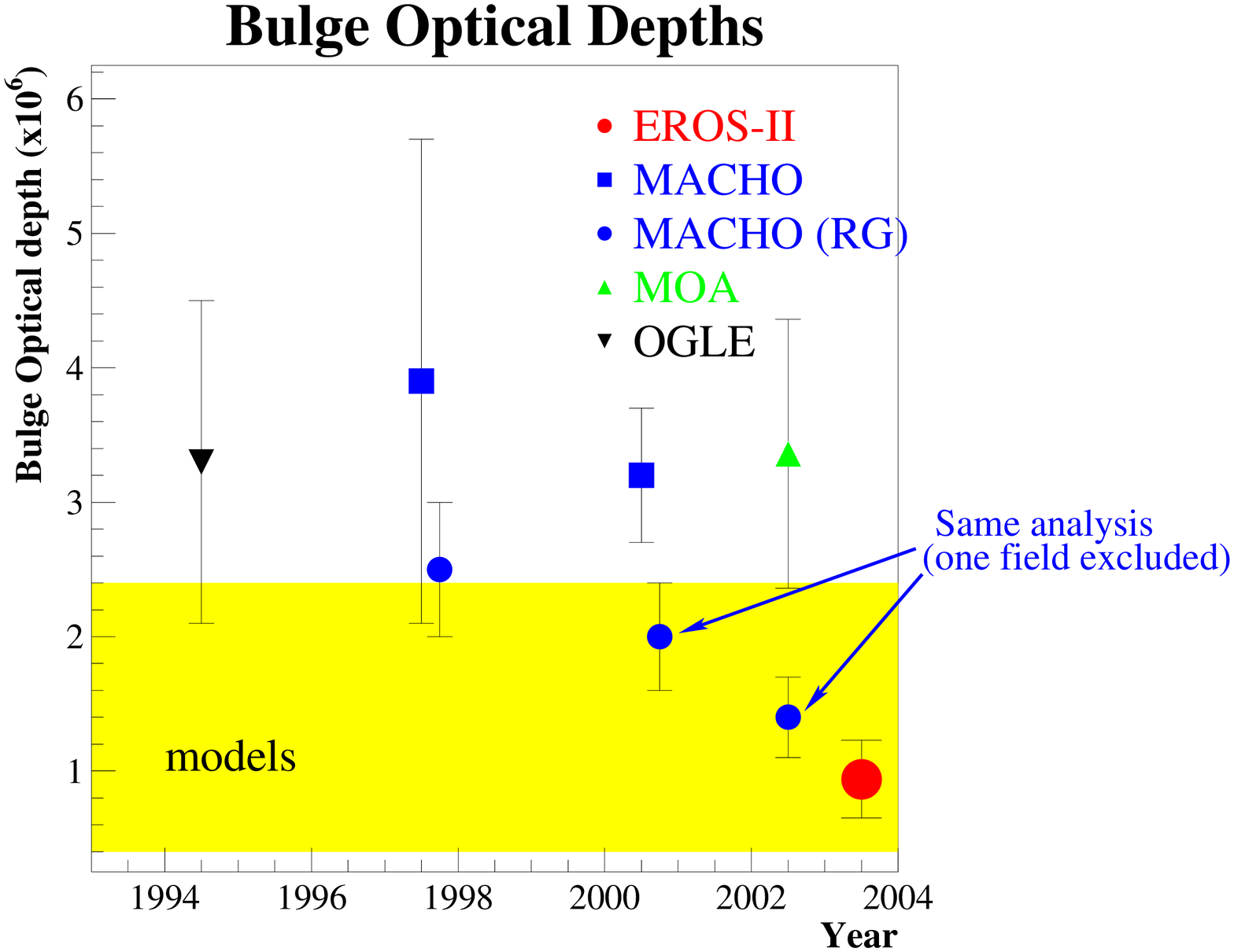}}  
\caption{Left: One of the EROS microlensing candidates toward the Galactic bulge.
Right: Measured bulge optical depths, from various groups}
\end{center}
\label{figbulge}
\end{figure}

\section{Conclusion}
The EROS2-MARLY instrument has been successfully operated in La Silla for seven years,
from July 1996 to February 2003, producing a huge amount of photometric data 
for microlensing and variable stars studies toward Magellanic clouds and in the Milky Way. 
The microlensing searches performed on less than the total available data 
rules out significant contribution of MACHOs in mass to the standard halo,
and constrains the mass distribution in the Galaxy.
However, although the observations has been stopped, the analysis of complete 
EROS data set should improve significantly these constraints.    

\end{document}